\documentclass[%
 reprint,
longbibliography,
 amsmath,amssymb,
 aps,
floatfix
]{revtex4-1}

\usepackage{tabularx,booktabs}
\usepackage{graphicx}
\usepackage{dcolumn}
\usepackage{bm}
\usepackage[colorlinks = true,
            linkcolor = blue,
            urlcolor  = blue,
            citecolor = blue,
            anchorcolor = blue]{hyperref}
\usepackage{siunitx}

\usepackage{xcolor}

\AtBeginDocument{%
  \heavyrulewidth=.08em
  \lightrulewidth=.05em
  \cmidrulewidth=.03em
  \belowrulesep=.65ex
  \belowbottomsep=0pt
  \aboverulesep=.4ex
  \abovetopsep=0pt
  \cmidrulesep=\doublerulesep
  \cmidrulekern=.5em
  \defaultaddspace=.5em
}

\newcommand{\srb}{$^{88}$Sr~}

\newcommand{\srfc}{$^1$S$_0$\,--\,$^1$P$_1$\,}

\newcommand{\srsc}{$^1$S$_0$\,--\,$^3$P$_1$\,}
\newcommand{\srclock}{$^1$S$_0$\,--\,$^3$P$_0$\,}

\begin{document}
 
\title{Optically loaded Strontium lattice clock with a single multi-wavelength reference cavity}

\author{Matteo Barbiero}
\author{Davide Calonico}
\author{Filippo Levi}
\author{Marco G. Tarallo}
\email{m.tarallo@inrim.it}
\affiliation{%
 Istituto Nazionale di Ricerca Metrologica, Strada delle Cacce 91, 10135 Torino, Italy
}%
\date{Last version:~\today}

\begin{abstract}
We report on the realization of a new compact strontium optical clock employing a {two-dimensional magneto-optical-trap (2D-MOT)} as cold atomic source and a multi-wavelength cavity as  frequency stabilization system. {All needed} optical frequencies are stabilized to a zero-thermal expansion high finesse optical resonator, and can be operated without frequency adjustments for weeks. We present the complete characterization of the apparatus. Optical control of the atomic source allows {us to perform} low-noise clock operation without atomic signal normalization. Long{-} and short-term stability tests of the clock have been performed for the \srb bosonic isotope  by means of interleaved clock operation. { Finally, we present the first p}reliminary accuracy budget {of the system}.
\end{abstract}

\maketitle

\section{Introduction}

{O}{ptical} lattice clocks (OLCs) based on neutral atoms are at the forefront of frequency metrology, exceeding current SI primary frequency standards of more than two orders of magnitude both in stability~\cite{Campbell2017,Schioppo2017} and accuracy~\cite{Beloy2021}. Therefore OLCs are suitable candidates for the future redefinition of the unit of time~\cite{Riehle2015}. 
Among several candidates, strontium is one of the most widespread atomic species in metrological and ultra-cold research laboratories. Because of its simple electronic structure and its commercial accessible cooling transitions, it finds successful application for optical clocks~\cite{Ushijima2015,Schwarz2020}, quantum control and quantum simulation \cite{Madjarov2019,Young2020}, probing new physics beyond the standard model \cite{Miyake2019} and chronometric geodesy \cite{Grotti2018,Takamoto2020,Bothwell2021}.

OLCs are essentially composed by three main parts: i) the optical local oscillator, with its local frequency reference usually consisting of an ultra-stable passive optical resonator: ii) the atomic frequency discriminator, which is a complex ultra-cold atomic apparatus to cool and trap atoms in tens of \SI{}{\micro K }-deep optical lattices; iii) and a self-referenced optical frequency comb for {frequency} measurement and comparison. 

Concerning the preparation of the atomic frequency discriminator, efficient laser cooling and trapping of ultra-cold atoms requires additional sub-systems for laser frequency stabilization and, in the case of the narrow \srsc Sr intercombination transition, spectral narrowing. This makes an OLC a rather complex system. Previous approaches to build compact and transportable OLCs tackled this problem by simplifying the frequency stabilization scheme either by employing a monolithic multicavity~\cite{Nevsky:13}, by sub-harmonic frequency dissemination of remote frequency references on a telecom network and subsequent optical phase-locked loops~\cite{Ohmae2021}, or by employing a single multi-color ultrastable cavity~\cite{Milani2017}. Furthermore, fast loading rates of the atomic sample are realized by employing Zeeman slowers or direct line-of-sight collimated atomic sources, whose collisions with trapped atoms limits both the trapping lifetime and the systematic uncertainty~\cite{Gibble13} if an in-vacuum shutter is not employed.

In this work we present a Sr optical atomic clock apparatus which simplifies the atomic frequency discriminator system. We employ an optically-controlled cold atomic beam source based on a sideband-enhanced two dimensional magneto-optical trap (2D-MOT)~\cite{Barbiero2020}, and a multi-wavelength frequency stabilization system, which also provides the short-term stability to the clock laser source. We study the stability and reliability of the apparatus by employing the bosonic \srb isotope, which possess the highest natural abundance, by means of the magnetic field-induced spectroscopy (MIS) method~\cite{Taich2006}.

The paper is organized as follows: we first describe the multi-wavelength frequency stabilization system (Sec.\ref{sec:freq}), its stability performances and its use for the clock laser stabilization. Then we present the atomic cooling and trapping apparatus (Sec.\ref{sec:man}) where we perform efficient two-stage magneto-optical trapping (MOT) and loading into a ``magic-wavelength'' optical lattice~\cite{Takamoto2005}. Finally, we show the results of the Sr optical lattice clock operating on the forbidden \srclock transition by the  magnetic-field induced spectroscopy method (Sec.\ref{sec:OL}).

\section{Compact multi-wavelength frequency stabilization system}\label{sec:freq}

\begin{figure}[tb]
    \centering
    \includegraphics[width=0.45\textwidth]{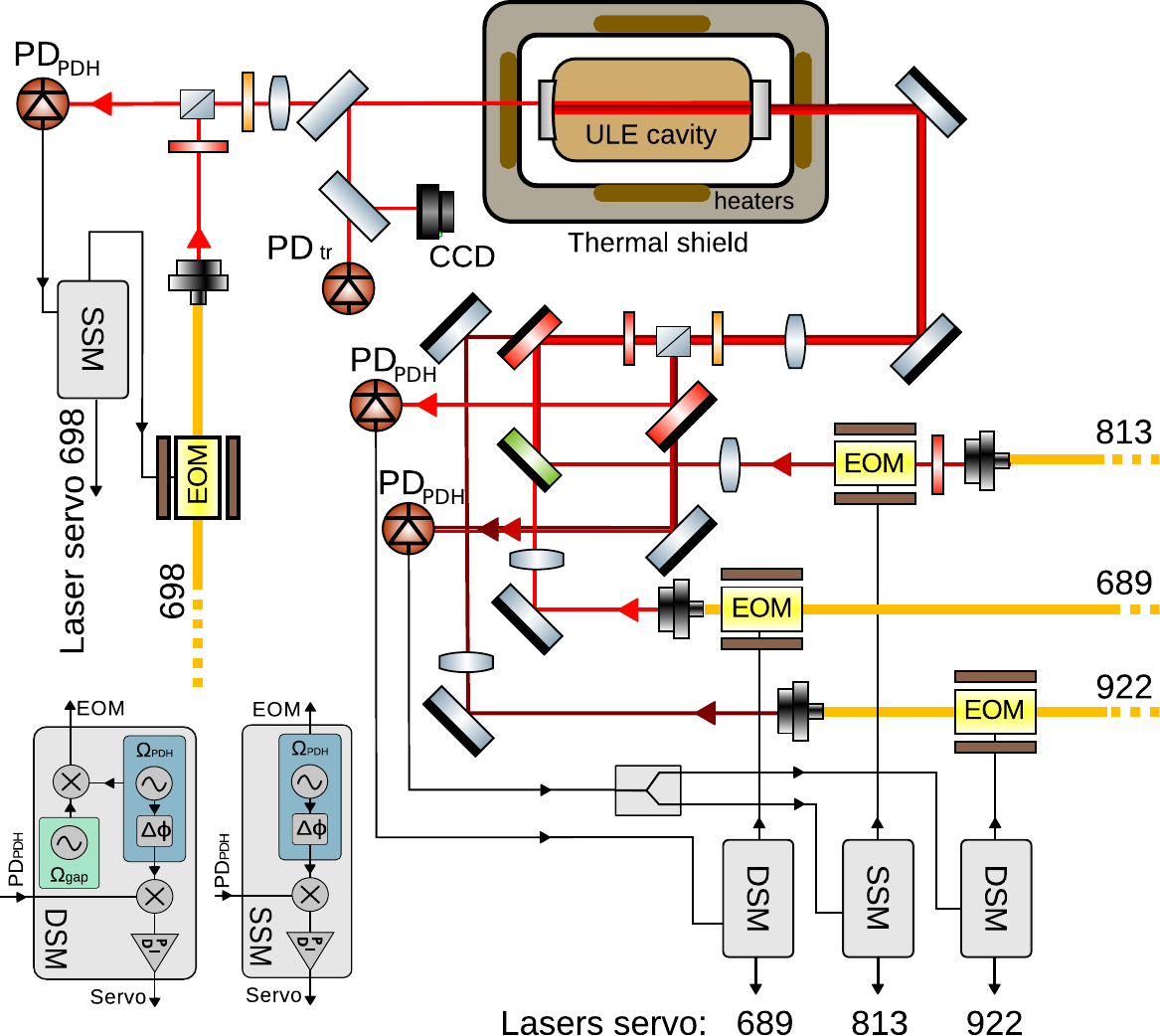}
    \caption{Schematic drawing of the multi-wavelength frequency stabilization system for a strontium optical lattice clock. {Red (green) mirrors represent long-(short-)pass optical filters.} PD: photodetector; SSM: single-sideband modulation/demodulation electronic circuit. DSM: dual sideband. Details in the main text.}
    \label{fig:TCCsetup}
\end{figure}

{Our} multi-wavelength frequency stabilization system is schematically shown in Fig.~\ref{fig:TCCsetup}. The {core} of the system consists of a monolithic, cylindrical optical cavity of length $L$ = 10 cm and diameter $\phi$ =  5 cm, made by ULE~\cite{Corning}. {The high-reflectivity coating of the ULE mirrors features three peaks at 922, 813 and 689 nm.} These correspond to the three main cooling and trapping wavelengths. {The highest design finesse, about \num{1.6(4)e4}, is reached for the narrow \srsc intercombination transition wavelength at 689 nm, while the other two wavelengths (first cooling stage and lattice trapping) have finesse  of \num{3.2(4)e3}.}

The cavity {sits} on a V-shaped support over 4 viton balls inside a stainless-steel vacuum tank. This{,} in turn{,} is thermally decoupled from the optical table by two teflon supports. High-vacuum (about \SI{1e-8}{mbar}) is mantained by a 20 L/s ion pump. The temperature of {the }cavity is actively stabilized to its zero-CTE point at \SI{29.34}{\degree  C} by means of one polymide thermofoil heater and two silicon rubber heaters attached to the three sides of the cylindrical-shaped vacuum system{. S}everal layers of polyurethane foam ensures thermal insulation. {A} digital proportional-integral-derivative servo loop keeps the vacuum tank temperature at the desired value within 5 mK.

Photothermal effects and residual temperature fluctuations can further limit the frequency stability of a laser locked to the optical cavity. These effects have been considered and described in the following sections.

\subsection{Cooling and trapping lasers frequency stabilization}

The two lasers needed for laser cooling of \srb at 461 nm and 689 nm are frequency referenced to the multi-wavelength cavity by means of the dual-sideband offset locking technique~\cite{Thorpe2008}. This modified version of the Pound-Drever-Hall (PDH) technique allows {us} to tune the carrier frequency independently from the cavity resonance by shifting the modulation frequency $\Omega_\text{gap}$. { The} PDH signal is extracted by demodulating the photodiode signal at the second sideband $\Omega_\text{PDH}$. Together with the lattice laser at 813 nm, all these three laser beams are sent to the same side of the cavity and share the same optical path and polarization optics, as shown in Fig.~\ref{fig:TCCsetup}. {T}he 689 nm {reflected} beam is separated from the other two {beams} by another short-pass mirror and then detected {with a} PDH photodetector. For the other two wavelengths, the error signal is generated from the photocurrent of the same photodiode by frequency demodulation at their respective $\Omega_\text{PDH}$ frequencies.

The \SI{689}{nm} laser for the narrow \srsc intercombination transition is a commercial extended-cavity diode laser (ECDL, Toptica DLPRO). It is partially sent to the multi-wavelength cavity through a polarization-maintaining (PM) optical fiber and phase modulated by a  fiber-coupled, wideband, electro-optic modulator (EOM, Jenoptik PM705).

The dual sideband modulation is generated by electronic mixing of two RF oscillators at $\Omega_\text{gap} \sim \SI{166}{MHz}$ and $\Omega_\text{PDH}=\SI{10.5}{MHz}$ respectively, so that the carrier has no phase modulation at $\Omega_\text{PDH}$. 
{Each offset sideband takes typically \SI{24}{\percent}  of the total power sent to the cavity, which is about \SI{45}{\micro W}. This corresponds to a modulation index nearly equal to \num{1}.} The correction frequency is then fed back to both the ECDL's piezotransducer for low frequency corrections{,} and to the diode current {modulation input} through a passive electrical network. Typical servo bandwidths of    \SI{600}{kHz} are achieved, so that the laser low-frequency instability is dominated by {the instability of the cavity}.

The \SI{461}{nm} light necessary for {the} \srfc transition is generated by a frequency-duplicated commercial diode laser (LEOS Solutions). We take a pick-off of the \SI{922}{nm} sub-harmonic decoupled from the main beam by an optical isolator and we send it to the multi-wavelength cavity from a PM fiber and a fiber-EOM. The offset sideband frequency is tuned to \SI{189.9}{MHz} while the second sideband is at \SI{13.33}{MHz}. About \SI{30}{\micro W} of optical power is sent to the cavity with a carrier-to-offset sideband power ratio of \SI{27}{\percent}. In this case the correction signal is fed back only to the ECDL's piezotransducer for slow correction of the seed wavelength.

The lattice laser (Ti:Sapph) at \SI{813}{nm} is frequency stabilized to the multi-wavelength cavity with standard PDH technique on the nearest cavity resonance to the known magic frequency~\cite{Akatsuka08}, with a stability exceeding \SI{100}{kHz}.

\subsection{Clock laser frequency stabilization}

The clock laser at \SI{698}{nm} is also a commercial ECDL (Toptica DLPRO) delivering up to \SI{35}{mW} of optical power. It is currently frequency stabilized to the multi-wavelength cavity, entering from the opposite side of the cavity with respect to the \SI{689}{nm} and the other cooling lasers, as shown in Fig.~\ref{fig:TCCsetup}. In order to avoid unwanted cross-talks with the close 689 nm laser light transmitted from the cavity, the input circular polarization is carefully tuned with opposite sign. We send \SI{20}{\micro W} of optical power phase modulated at $\Omega_\text{PDH} = \SI{23}{MHz}$, far from any harmonics of the other RF frequencies.

The clock laser is locked to cavity with the standard PDH technique. The servo loop is similar to the one described for the \SI{689}{nm} laser, with correction sent to both {the }diode current and the ECDL piezotransducer. In this case the control bandwidth exceeds \SI{1.2}{MHz}, with the in-loop error signal reaching the detector noise floor (\SI{85}{mHz/\sqrt{Hz}}) up to 20 kHz.

\subsection{Stability results}

\begin{figure}[tb]
    \centering
    \includegraphics[width=0.45\textwidth]{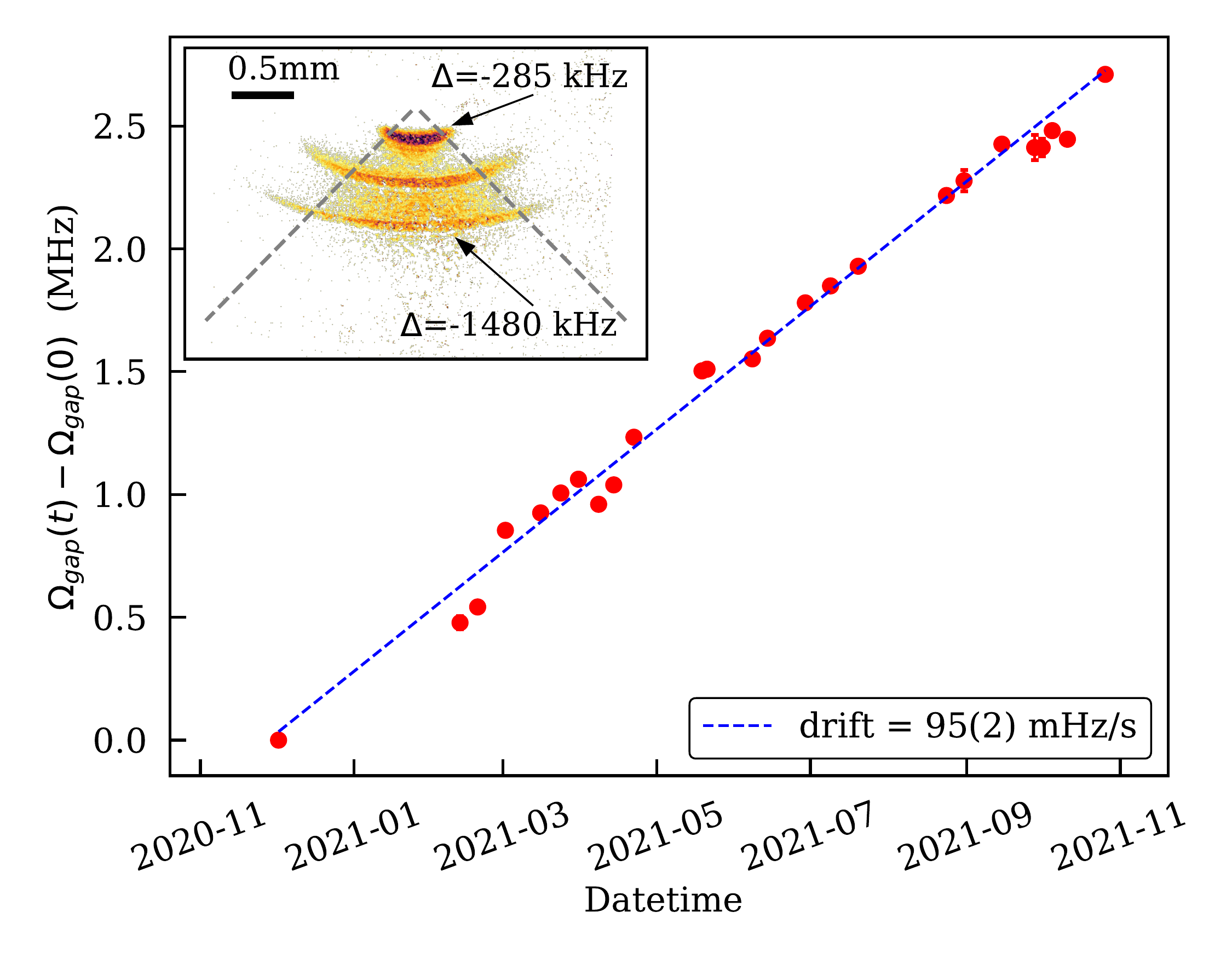}
    \caption{Long-term frequency stability of the multi-wavelength cavity measured for a 11 months period by monitoring of the offset sideband $\Omega_\text{gap}$ employed in the frequency stabilization loop of the \srsc transition laser.  In the inset, absorption images of the Red MOT showing transverse dimension dependence at different detunings $\Delta$. }
    \label{fig:merda1}
\end{figure}

The frequency stability of an optical resonator of ULE, which is temperature-stabilized to its zero-coefficient of thermal expansion (CTE) point, is limited by the aging of the spacer material, resulting is a slow drift~\cite{Haefner2015}. We developed a method to track this long-term drift with respect to the resonance transition \srsc necessary for the second-stage ultra-cold MOT of the atomic sample (see Sec.~\ref{sec:man}). We infer the \srsc resonant frequency by imaging the red MOT with respect to the offset sideband frequency $\Omega_\text{gap}$ and then fitting the transverse dimension of the atomic cloud{,} which linearly depends on the cooling frequency detuning~\cite{Katori1999} as shown in the inset of Fig.\ref{fig:merda1}.  The main panel of  Fig.~\ref{fig:merda1} reports the recorded values of $\Omega_\text{gap}$ at which the 689 nm laser is on resonance with respect to the \srsc transition over a period of 11 months. We infer an average drift rate of $\SI{95(2)}{mHz \per s}$ (or \SI{2e-11}{\per day} in relative units). This low drift value allows us to prepare and manipulate ultra-cold samples of Sr atoms without any frequency adjustment for weeks.

The stability of the multi-wavelength cavity as frequency reference is limited by vibration-induced length fluctuations $\delta L_v$ induced by axial and transverse cavity accelerations $a_\text{axial}$ and $a_\text{trans}$. This can be parametrized as~\cite{Hall1999}

\begin{equation}
\frac{\delta L_v}{L} = \frac{\rho}{2Y}\left(\epsilon L a_\text{axial}+a_\text{trans} \phi\sigma\right),
\end{equation}
where $\rho$ is the cavity spacer density, $\sigma$ and $Y$ are the Poisson ratio and Young modulus respectively, and $\epsilon$ parametrizes the acceleration transmissivity due to the cavity support geometry, ranging from {0 to 1~\cite{Nazarova_2006}}. No antivibration measures are taken for our cavity, so that for typical ambient seismic noise we can estimate the induced frequency instability. Assuming a typical environmental acceleration noise spectral density $\delta a(f) \sim$ \SI{5e-5}{ms^{-2}/\sqrt{Hz}}~\cite{Fiori2004,Tarallo2011} up to few hundred Hz, and a sensitivity coefficient $\epsilon$ = 0.5, the expected vibrations-limited fast linewidth {is expected to be} about 10 Hz.

An important detrimental effect that could limit the frequency stability of a laser, and in particular the clock laser, to the multi-wavelength cavity is the length fluctuations due to other lights' intensity noise. While we do not expect that this effect limits our clock laser, due to the non-compensated vibration noise, it is interesting to evaluate this effect for future implementations of the system with higher frequency stability requirements. To estimate the transmitted intensity noise from the other lasers to the clock light, we modulated the amplitude of the RF power generating the offset sidebands $\Omega_\text{gap}$  for the \SI{689}{nm} and \SI{922}{nm} lights and we looked at the corresponding frequency shift with respect to a high-stability optical oscillator~\cite{Barbiero2021}. From a linear fit of the cavity frequency shifts against the transmitted power, we infer {cavity shift coefficients} of { $k_\text{689}=\SI{4(1)}{Hz/\micro W}$} and $k_\text{922}=\SI{2(2)}{Hz/\micro W}$ respectively. The shift measurement for the 922 nm light is compatible with zero. These results can be explained as due to the heating of the dielectric mirror coating   generated by the intracavity optical power at different wavelengths ~\cite{Bergquist1992,Tarallo2011}. For our system, this results in a frequency instability $\sigma_y(\tau) = \SI{1.1(4)e-16}{\sqrt{\tau}}$ for an intra-cavity relative intensity noise of { \SI{0.3}{\percent}}.

The short-term stability of the multi-wavelength frequency stabilization system is best studied for clock laser by looking at its effect on the clock spectroscopy and the clock stability via the Dick effect~\cite{Quessada_2003}. This is described in detail in Sec.~\ref{sec:stability}.

\section{Cooling, trapping and probing apparatus}\label{sec:man}

\begin{figure}[!t]
    \centering
    \includegraphics[width=0.45\textwidth]{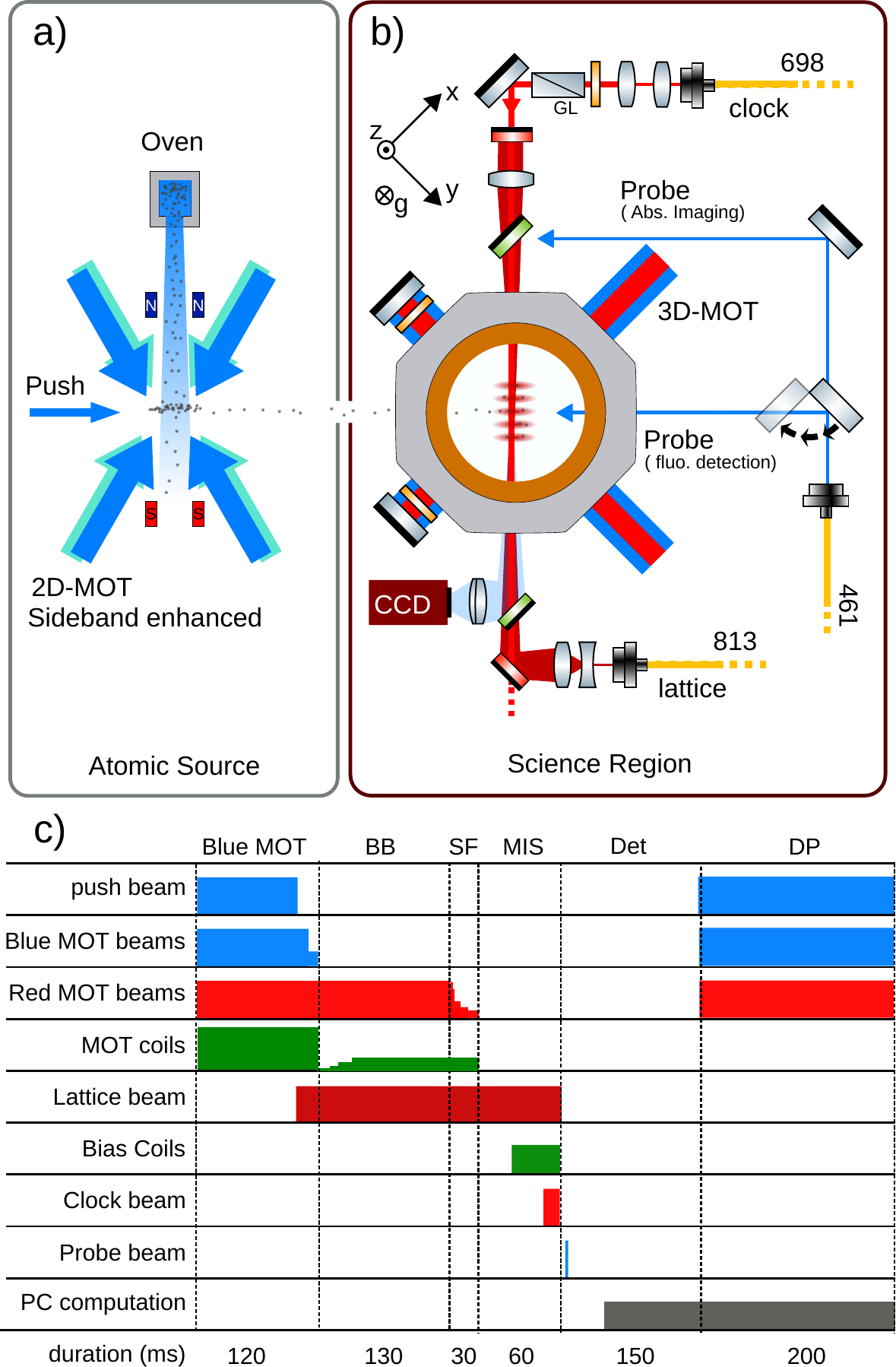}
    \caption{Schematic drawing of optical lattice clock system and its experimental procedure. \textbf{a)} Atomic source region,  \textbf{b)} science cell  and schematic layout for  atomic interrogation, \textbf{c)}   clock interrogation procedure. (BB) Broadband Red MOT phase, (SF) Single Frequency Red MOT phase, (MIS) Magnetic induced spectroscopy, (Det.) Detection, (DP) Data processing }
    \label{fig:fig1_app}
\end{figure}

A schematic overview of our atomic cooling and trapping apparatus is shown in Fig.~\ref{fig:fig1_app}. It consists of two main parts, the 2D-{MOT} atomic source and the science cell setup. 

The {c}old atomic source of this apparatus, as well as the details on the complete vacuum system, has been extensively described in~\cite{Barbiero2020}. Here we briefly recall the main properties of the system. A cold, bright atomic beam is generated by a two-frequency {2D-MOT} transversely loaded from a collimated Sr oven (typically operated at \SI{460}{\celsius}), with an average longitudinal velocity of \SI{22}{m/s} and a transverse temperature of less than \SI{1}{mK}. This allows {us} to achieve a loading rate  in the science chamber  up to \SI{8e8}{atoms \per s}, tunable by changing the optical power of the push beam.  

Such cold atomic beam is then cooled and trapped in the science cell, where three-dimensional magneto-optical cooling and trapping is performed before loading {t}he atomic ensemble into the optical lattice for clock spectroscopy. The overall procedure  for probing the \srclock transition by mean{s} of MIS technique is  depicted in the Fig.~\ref{fig:fig1_app}(c).

\subsection{Laser cooling of $^{88}$Sr atoms}

In the science chamber,  the atoms from the atomic source are loaded in the ``Blue MOT'' operated on the \srfc at \SI{461}{nm} for \SI{100}{ms}. We typically apply a total intensity $s=0.88$ (in units of the resonant saturation intensity) and a magnetic field gradient of \SI{4}{mT/cm}.  {Switching off the push beam interrupts the atomic flux. The MOT is operated for additional 20 ms. This covers both the time-of-flight of the remaining atoms from the 2D-MOT, and a short MOT phase (\SI{5}{ms}) at reduced intensity ($s=0.26$) to further cool down the collected sample. At the end we collect  up to \SI{7e6}{atoms} at \SI{2(1)}{mK}.} 
All {the} laser beams at 461 nm and the magnetic field gradient are then switched off, {and we} use a mechanical shutter to completely turn off the blue light{.}

The second cooling stage is performed on the  { \srsc }  intercombination transition (``Red'' MOT~\cite{Katori1999}) at 689 nm. We initially employ about \SI{13}{mW} {of }optical power ($s \sim \num{2e3}$) with a Broad-Band (BB) spectrum to cover the majority of the Doppler spectrum of the atoms released from the Blue MOT. A double-pass acousto-optic modulator (AOM) yields the broadened spectrum with an FM frequency of \SI{35}{k Hz}, modulation depth of \SI{3}{MHz} and a minimum detuning of \SI{-400}{kHz} from the atomic resonance. During the BB Red MOT phase, the magnetic field gradient is ramped up to \SI{1.3}{mT/ cm} in \SI{26}{ms}. The atomic density and cloud dimensions reach their stationary values in \SI{130}{ms}. 
At the end of the BB phase, we trap up to \SI{4e6}{atoms} at \SI{11}{\micro K}.

Finally, the temperature of the atomic sample is further reduced by a Single-Frequency (SF) red MOT phase. Here the optical power is exponentially ramped down from \SI{13}{mW} to \SI{10}{\micro W} ($s=\num{10}$) in \SI{30}{ms}. At the end of this cooling stage, we trap \SI{3.5e6}{atoms} below \SI{1}{\micro K}. Also for the 689 nm light, a mechanical shutter is used to avoid residual stray light going to {the} atoms in the optical lattice.

During the whole Red MOT{,} the optical lattice beam is turned on, so that {the atoms overlapped to the lattice beam remain trapped and ready for clock spectroscopy at the switching off of the red beams and the magnetic field gradient}. The total duration of the cooling and trapping is {about} 300 ms.

\subsection{Magic-wavelength optical lattice}

\begin{figure}[!t]
    \centering
    \includegraphics[width=0.49\textwidth]{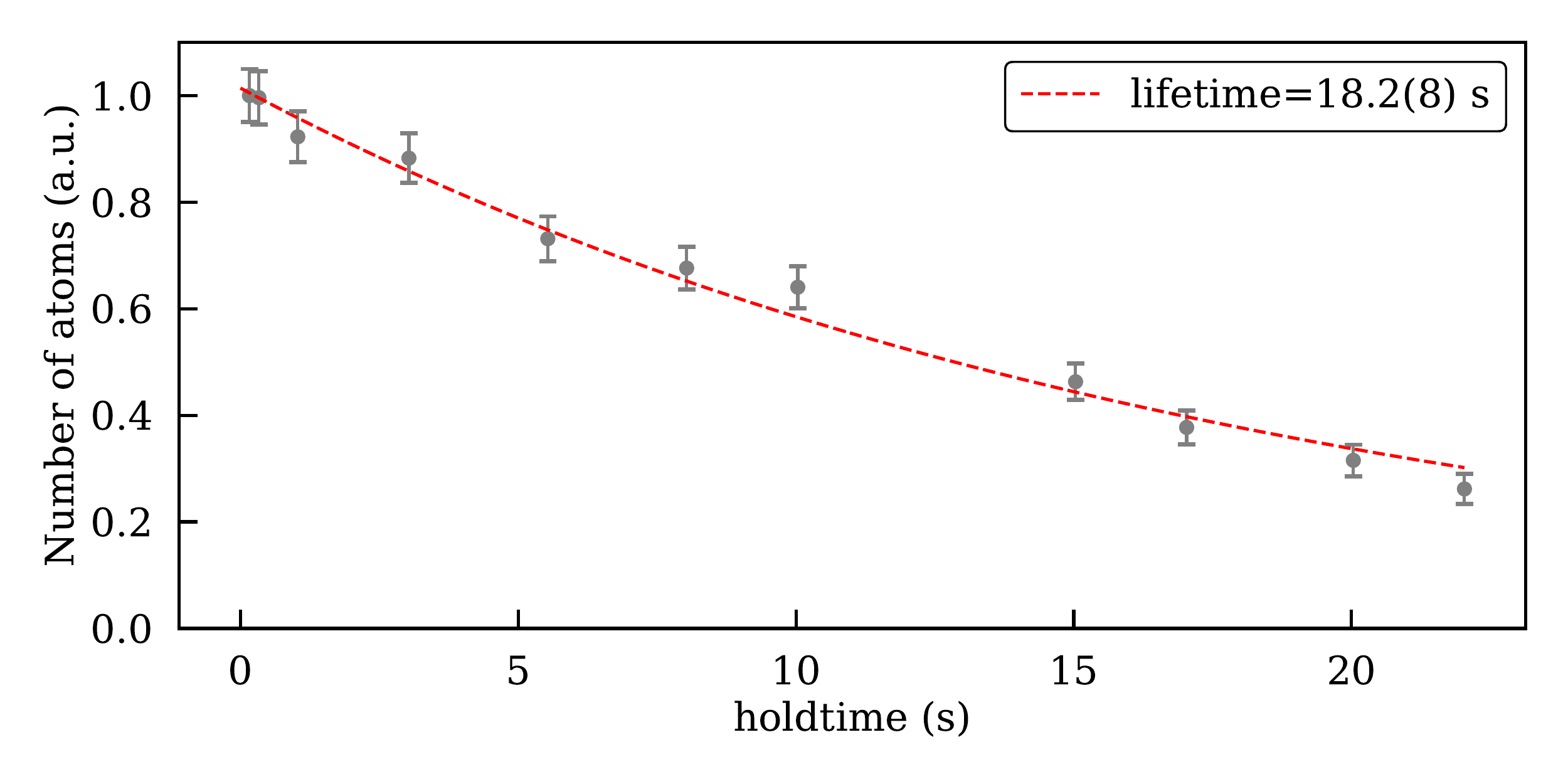}
    \caption{Numbers of atoms in the optical lattice as a function of different {hold time}. The number of atoms are normalized to their maximum value.}
    \label{fig:lattice_lifetime}
\end{figure}

The optical lattice trap is {realized with} \SI{500}{mW} of optical power at the magic wavelength of \SI{813}{nm}, { delivered with} a PM optical fiber. The beam is shaped to produce a beam waist of about \SI{50}{\micro m} in correspondence of the red MOT center. The lattice axis is orthogonal to the gravity direction as shown in {Fig.}\ref{fig:fig1_app}. The lattice retro-reflection dichroic mirror is carefully aligned maximizing the amount of power coupled back to {the} fiber collimator.   Up to \SI{4e5}{atoms} can be loaded in the lattice trap with a measured lifetime of \SI{18.2(8)}{s}, as plotted in Fig.~\ref{fig:lattice_lifetime}. This value is dominated by the off-resonant scattering ($\Gamma_{sc}^{-1} \approx \SI{33}{s}$ \cite{Grimm1999}), therefore vacuum-limited lifetime can be as high as \SI{40}{s}.
This exceptionally long lattice lifetime is a consequence of {the} 2D MOT loading that {allows us} to completely turn off the atomic flux during lattice spectroscopy{, and also maintaining} much higher differential vacuum between the two chambers  in comparison to a classic Zeeman slowe{r~}\cite{Drscher2018}. 

Time-of-{flight} absorption imaging measurements of the atomic sample show temperatures lower that \SI{2}{\micro K} in both the transverse directions, while the axial free-expansion of the atomic cloud is parallel to the imaging beam{,} thus the axial temperature {is} not measurable with this method. The transverse size of the optical lattice can be extracted from temperature measurements~\cite{Grimm1999}, and it is about $\sigma_r = $ \SI{14}{\micro m}. The axial dimension is not accessible by the imaging system, and it is roughly estimated equal to twice the Red MOT radius, about \SI{400}{\micro m}.

The optically-controlled atomic source combined with the cooling and trapping process yields a lattice population fluctuation as low as \SI{3}{\percent} for maximum loading rate, which nearly doubles at low loading rates. This number fluctuation is nearly half {of the one} observed when the 461 nm laser was frequency stabilized to an atomic vapor~\cite{Barbiero2020}.

\subsection{Magnetic field-induced spectroscopy setup}

Magnetic field-induced spectroscopy is enabled by an homogeneous bias magnetic field  applied by inverting the current of one of the MOT coils, thus switching from {a}nti-Helmholtz to {H}elmholtz configuration. The bias magnetic field {is} directed along the gravity direction and orthogonal to the lattice beam propagation direction.

The clock laser beam at 698 nm is directed towards the atoms from the dichroic retro-reflection lattice mirror (high-reflective at 813 nm and high-transmissive, $\sim\SI{90}{\percent}$, at the clock wavelength). The clock laser beam is shaped to have a waist of \SI{150}{\micro m}, i.e. three times the lattice beam waist, to ensure high homogeneity on the atoms. Clock light standing waves are avoided by using a second dichroic mirror along the optical path of the input lattice beam, as depicted in Fig.~\ref{fig:fig1_app}(b). We carefully align the clock beam to the atoms by  maximizing the amount of power injected to the fiber collimator of the lattice beam. Linear polarization parallel to the magnetic field is ensured by a Glan-Thomson polarizer placed before the lattice retro-reflection mirror. Finally, {power and duration of the clock laser pulse are controlled}  by an AOM before the input fiber.

A second AOM driven by an externally-referenced RF oscillator {(working around 345 MHz),} is used to tune the  frequency of the clock laser for clock spectroscopy scans and frequency stabilization by FM modulation. The RF oscillator is directly controlled by an analog output generated by the experiment control system~\cite{Barbiero2019}.

\subsection{Atomic detection setup}

Atomic detection and diagnostics is performed by the imaging system. It consists of a CCD camera (Stingray F-201, 1624 $\times$ 1234 pixels and \SI{4.4}{\micro m} pixel size) and an achromatic lens $f=\SI{100}{mm}$ displaced to get a numerical aperture of \num{0.10(4)} and magnification factor of \num{0.617(3)}. A probe beam{,} resonant with the \srfc strong transition{,} can be either sent to the CCD camera for absorption imaging of the atomic sample, or directed orthogonally to the absorption axis for fluorescence detection by means of a removable mirror. The absorption imaging optical path is integrated along the lattice direction by means of two dichroic mirrors. Residual light from the lattice and clock laser beams are then blocked by an interference filter peaked at \SI{461}{nm}{,} mounted in front of the CCD camera.

While absorption imaging is best suited for the atomic sample diagnostics (atomic cloud dimension, temperature and calibrated atomic count), we perform fast fluorescence imaging for clock spectroscopy. {In this case, the} probe beam pulse {has} a duration of \SI{0.7}{ms} and optical power of \SI{3}{mW}{, while its} linear polarization maximizes the atomic fluorescence towards the CCD. The  CCD camera exposure time is \SI{220}{\micro s}, while the image is downloaded to the {computer control} in less than \SI{50}{ms}. Only {\SI{3}{\percent}} of the total CCD array is employed to speed-up the data download and processing. Within the same image we define two regions of interest (ROIs) with equal areas. One of the ROIs covers the majority of the atomic cloud and provide{s} the photon counts, the other is placed at the corner of the image and provides the background signal of the CCD. The difference between the ROIs counts provides the fluorescence counts. The detected fluorescence signal is maximized by releasing the atoms from the lattice and probing the atoms after \SI{4}{ms} of free fall.

Almost \SI{200}{ms} are spent by the {computer control} to manage the CCD data processing and to generate the feedback signal to keep the system on the resonance of the \srclock transition. This represents the current limit for the duration of the experimental cycle. To further shorten the clock cycle, we are planning to implement a parallel processing of the CCD data within the preparation time of the atoms in the optical lattice trap.

\section{Results}\label{sec:OL}

\subsection{Resolved sideband spectroscopy}\label{sec:SRS}

\begin{figure}[!tb]
    \centering
    \includegraphics[width=0.49\textwidth]{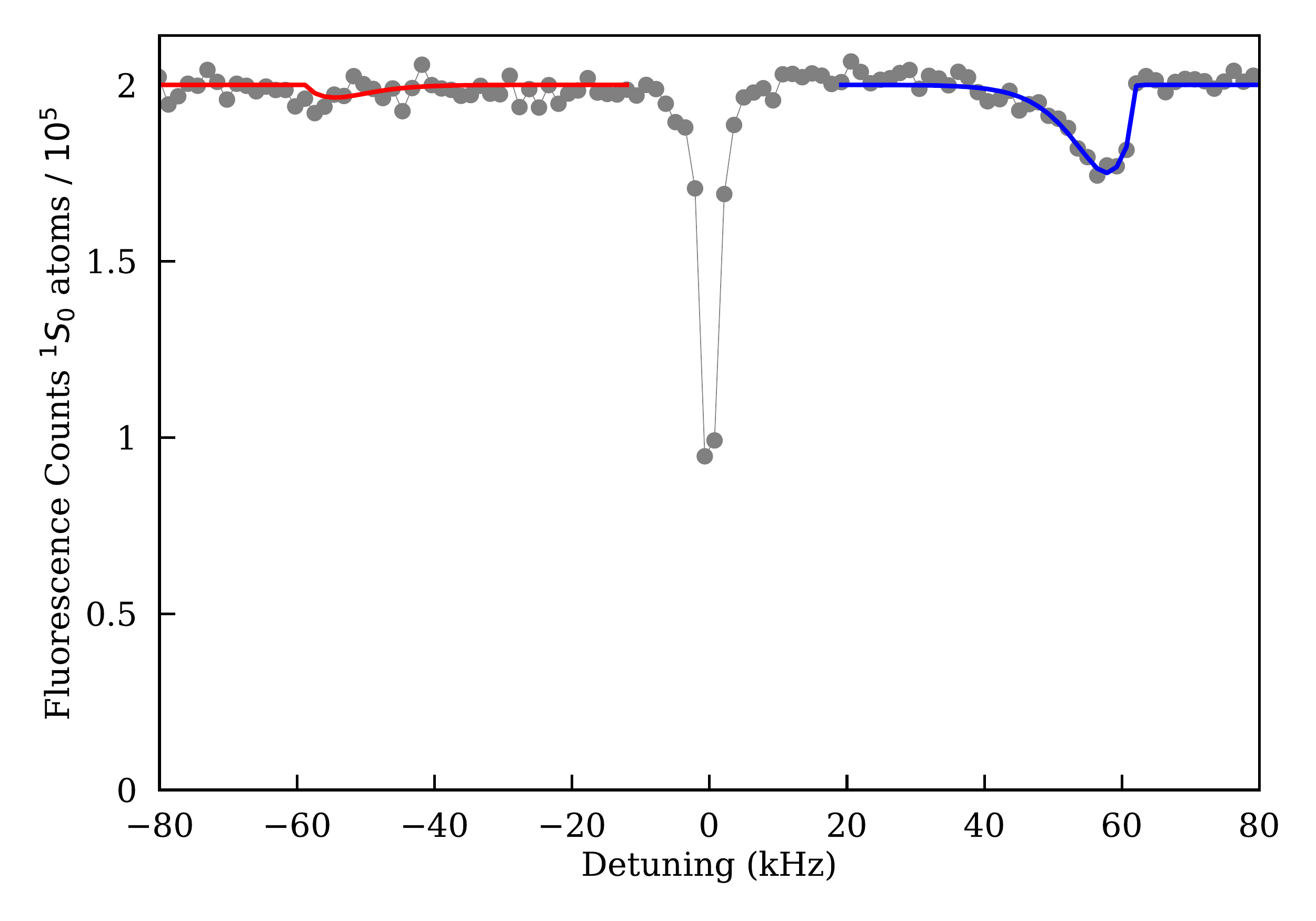}
    \caption{Sideband-resolved spectroscopy of the \srb clock 
transition. The trace is the average over four consecutive frequency scans. Motional sidebands (blue and red curves) are simultaneously fitted~\cite{Blatt2009} to estimate the trap depth and the apparent radial and axial temperatures.}
    \label{fig:SRS}
\end{figure}

We perform high resolution MIS of the \srb clock transition \srclock in the Lamb-Dicke regime with  motion  resolved  from  the  carrier. An example of the sideband spectrum is presented in Fig.~\ref{fig:SRS}. Here the bias coils are driven at maximum current $I_\text{coil}=\SI{10}{A}$ and a pulse duration of \SI{60}{ms} {is employed }to maximize the excitation of the motional sidebands. With a typical lattice optical power of \SI{450}{mW}, we measure an axial trapping frequency of \SI{65.5(2)}{\kilo Hz}, implying a lattice  depth   $U_0= 92.7(5) E_r$ = \SI{15}{\micro K},  where  $E_r$  =  $h^2/(2m\lambda_L^2)$ is  the  lattice  recoil  energy. The Lamb-Dicke parameter $\eta$ associated to our lattice trap depth is $\eta = 0.26$.

We also measure the apparent axial ($T_z$) and radial ($T_r$) temperatures of the atomic sample from the shape and the relative areas of the first-order motional sidebands~\cite{Blatt2009}. The resulting temperatures are $T_r = \SI{3.2(3)}{\micro K}$ and $T_z= \SI{1.6(3)}{\micro K}$. Compared to time-of-flight temperature measurement, the apparent radial temperature is slightly higher ($\sim \SI{2}{\micro K}$), implying either the simplicity of the {fitting function} which does not include any broadening {effect}, or an underlying heating mechanism due to photon-assisted collisions. 

\subsection{Narrow-line spectroscopy and clock stability test}\label{sec:stability}

High-resolution Rabi spectroscopy is realized by exciting the clock transition at the actual $\pi$ pulse for each configuration of magnetic field and probe intensity. The resulting linewidth can be narrowed by decreasing either (or both) the bias magnetic field {or} the probe power. Furthermore, because of collisional dephasing, narrow lines have to be obtained by lowering the lattice site density {r}educing the number of loaded atoms. 

\begin{figure}[tb]
    \centering
    \includegraphics[width=0.47\textwidth]{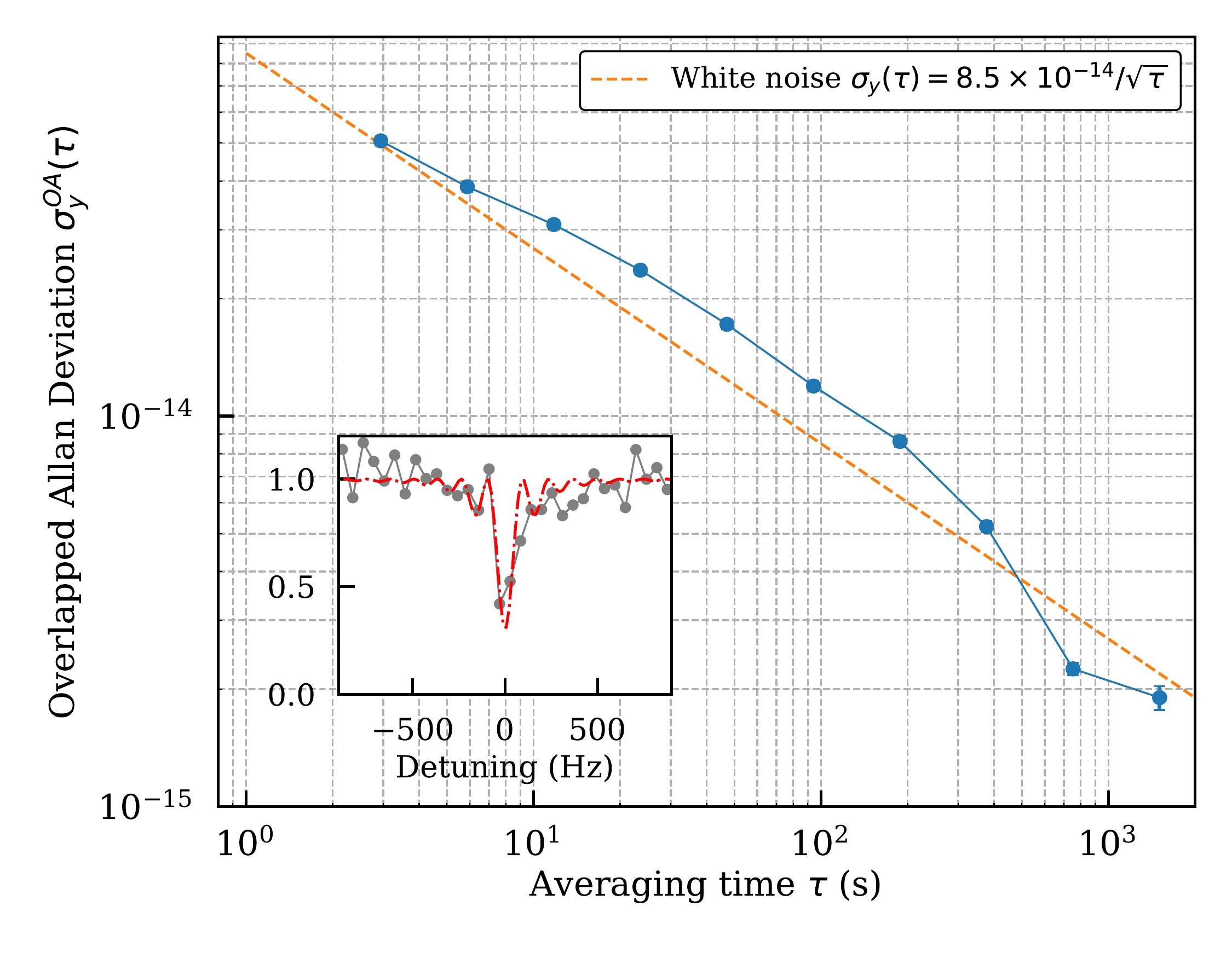}
    \caption{Main panel: Allan deviation of the difference between two interleaved stabilizations { at $N_{at}$ = \num{1.5e5} and \num{3e5}. The dataset covers a time span of 62 minutes.} 
    Inset panel: Narrow-line spectroscopy of the \srclock clock transition fitted with a Rabi response function. The corresponding FWHM $\approx$ 1.067 $\Omega$ = 86(7) Hz. }
    \label{fig:narrow}
\end{figure}

The inset of Fig.~\ref{fig:narrow} shows a narrow-line spectrum obtained by sweeping the clock light frequency for a 3 kHz span with bias coil current $I_\text{coil}$ = 3 A and clock power of 2.3 mW and with $2\times10^4$ atoms in the lattice for 8 ms. The resulting resonance has a full-width half-maximum of \SI{86(7)}{Hz} and shows   a \SI{70}{\percent}  contrast.

Clock operation is enabled by locking the clock laser frequency to an absorption feature, by means of two sequential pulses {separated in frequency} by an amount $f_{FSK}$. The frequency-shifting key (FSK) modulation depth is chosen to roughly match the FWHM of the atomic resonance. The stability of the clock operation is studied by interleaving between two values of some clock parameters. { Clock operations were carried on for several hours without unlocks or severe glitches. We have not observed any effect related to the long-term drift of our multi-wavelength cavity which would degrade the number of atoms loaded in the lattice.} A typical interleaved stability in fractional units is shown in Fig.~\ref{fig:narrow}.  {The resulting o}verlapping Allan deviation decreases asymptotically as  $\sigma_y^{OA}(\tau)=\num{8.5e-14}\ /\sqrt{\tau}$. This means that the single-operated clock, i.e. with half averaging cycle time {measurement noise}, has an average asymptotic stability of $\num{4.2e-14}/\sqrt{\tau}$. 

The white-noise-limited clock instability can be due to the local oscillator instability via {the} Dick effect~\cite{Quessada_2003}, or by shot-to-shot fluctuations of the non-normalized number of lattice atoms. The former should be particularly limiting for our clock because of the low duty cycle (about 1.3\% for a Rabi-limited resonance and 10 ms pulse). {If} we assume a Flicker noise floor similar to that of Ref.~\cite{Pizzocaro2012} and a white-noise of \SI{7.3}{Hz^2/Hz} limited by vibration noise{,} we get a Dick-limited Allan deviation of \num{4e-14} at \SI{1}{s}, nearly as much as the measured data.

\subsection{Clock frequency shifts}

\begin{figure*}
    \centering
    \includegraphics[width=0.44\textwidth]{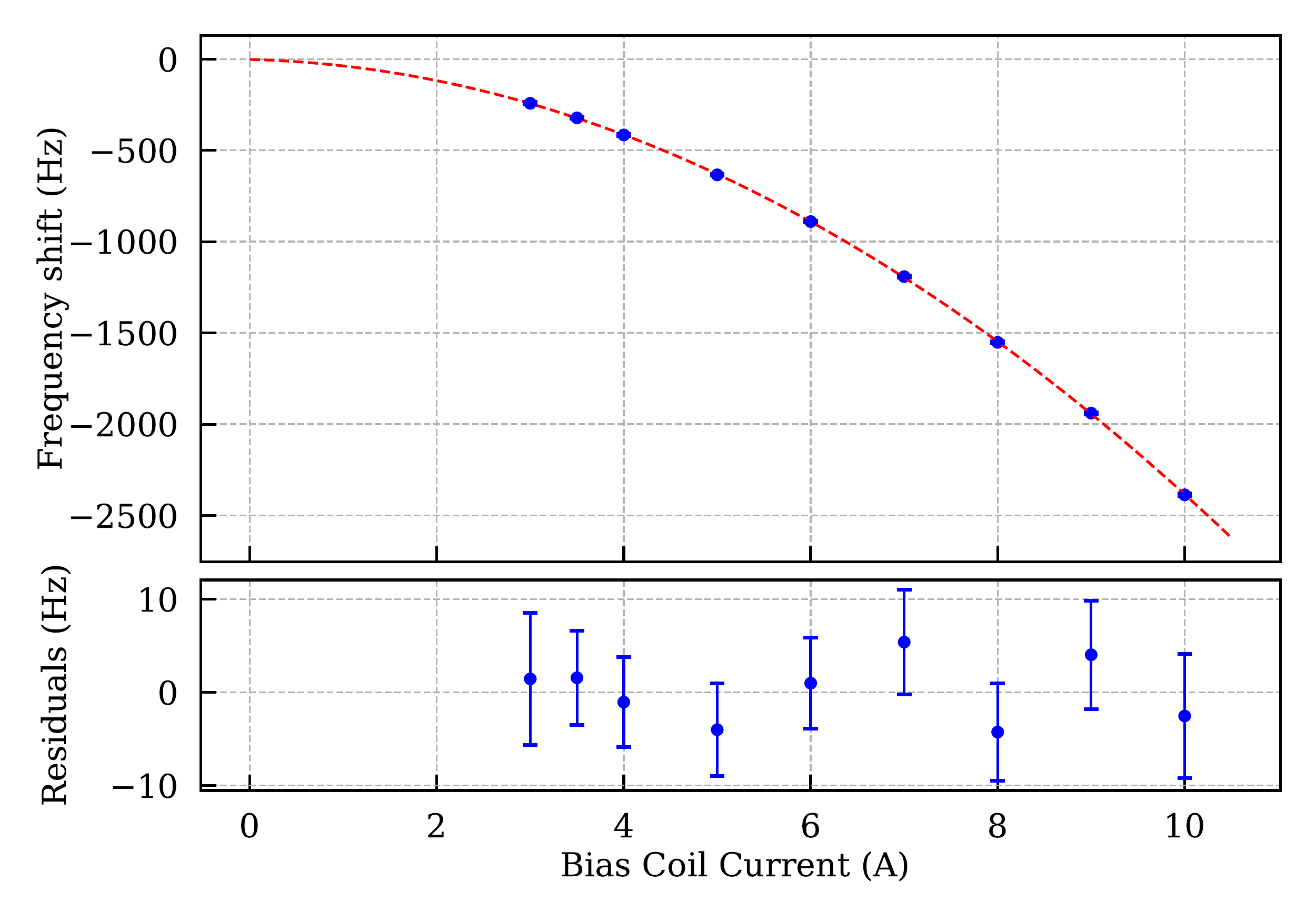}
    \includegraphics[width=0.44\textwidth]{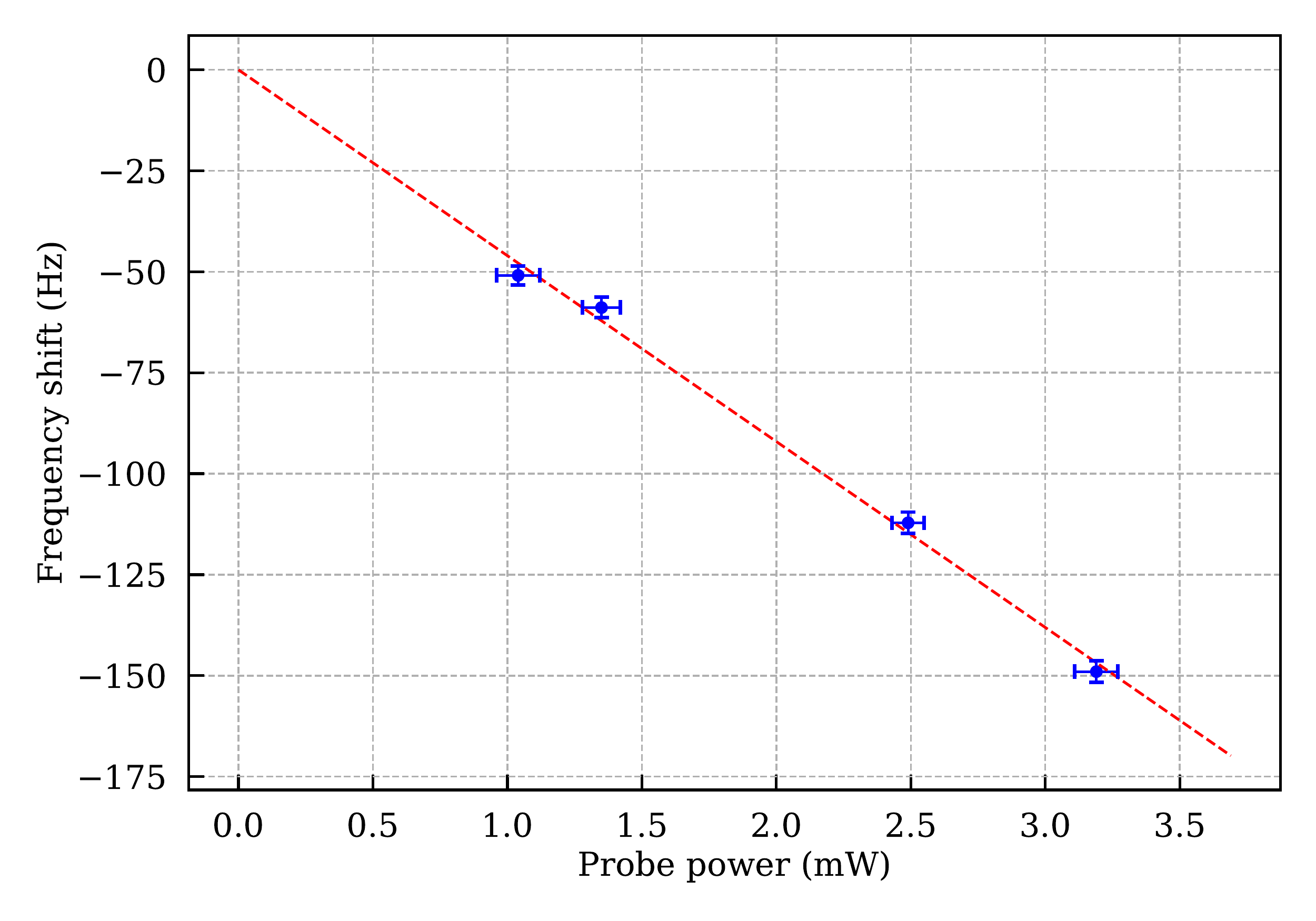}
    \caption{Systematic study of MIS clock frequency shifts. The absolute values of both the magnetic field and the probe light intensity were calculated using coefficients from~\cite{Taich2006}. On the left panel, measurement of the second-order Zeeman shift as a function of the bias coil current together with fit residuals. On the right panel, evaluation of the probe light shift as function of the input power. }
    \label{fig:clock_calibration}
\end{figure*}

{Understanding} the main sources of uncertainty {is} necessary to calibrate and test the stability of our apparatus. We have performed a preliminary evaluation of the systematic effects in our \srb optical clock by interleaved frequency measurements. The typical measurement duration is 15 minutes for each point. The main {sources} of systematic  shifts  and their uncertainties are summarized in Table~\ref{tab:budget}.

\begin{table}[tb]
    \caption{\label{tab:budget}Preliminary accuracy budget for the typical experimental conditions of our \srb clock (see main text for details). All reported values are in Hz.}
    \centering
    \begin{tabular}{lcc}
        \toprule
        Effect & Shift & Uncertainty\\
        \midrule
         AC Zeeman & -243.7 & 3.7\\
         Probe light &-105.8& 5.3\\
         Lattice light & 0 & 5.4\\
         Density & 1.9& 0.2\\
         BBR & -1.90 & 0.01 \\
         Frequency chain & 0 & 0.02\\
        \midrule
          \textbf{Total:} & \textbf{-349.6} & \textbf{8.4}\\
         \bottomrule
    \end{tabular}
\end{table}

Because of the artificial coupling of the two clock levels in the bosonic clock, the two  most  important  contributions  to  the uncertainty budget are the quadratic Zeeman shift and the light shift from the \SI{698}{nm} clock laser. {The associated} shift coefficient are very well known for {Sr} both theoretically and experimentally~\cite{Taich2006}, thus we use these values to calibrate the bias magnetic field and the probe intensity. The results are shown in Fig.~\ref{fig:clock_calibration}. The quadratic Zeeman shift $\Delta\nu_{B}$ {can be expressed as a } quadratic function of the bias coils current with an offset $B_0$, $\Delta\nu_B = \beta (k_\text{coil} I_\text{coil}+B_0)^2$, with {  $\beta = \SI{-23.8(3)}{Hz/mT^2}$}  \cite{Nicholson2015}. The measured bias coils  current calibration coefficient is $k_\text{coil}$ = 0.972(7) mT/A, so that for a typical value of the bias coils current ($I_\text{coil}$ = 3 A), the quadratic Zeeman shift induced by {a} bias field of \SI{2.9}{mT} is resolved with an  uncertainty of 3.7 Hz. The probe light shift is calibrated with a linear function resulting in a frequency shift of  { \SI{-46(2)}{Hz/mW}}, implying a probe beam width of \SI{158(4)}{\micro m}. Thus, we can express the effective Rabi frequency as function of the bias coil current and probe power by means of the respective induced shifts as~\cite{Taich2006}

\begin{equation}
\Omega_R = \xi\ I_\text{coil}\sqrt{P_L}\quad ,
\end{equation}
where $\xi \approx $ \SI{10.6}{Hz/(A\sqrt{mW})}.

The scalar light shift from the 813 nm lattice laser was estimated by interleaving different values of its power. The measured uncertainty at typical working lattice depth is about 5.4 Hz due to statistical uncertainty.

\begin{figure}[tb]
    \centering
    \includegraphics[width=0.45\textwidth]{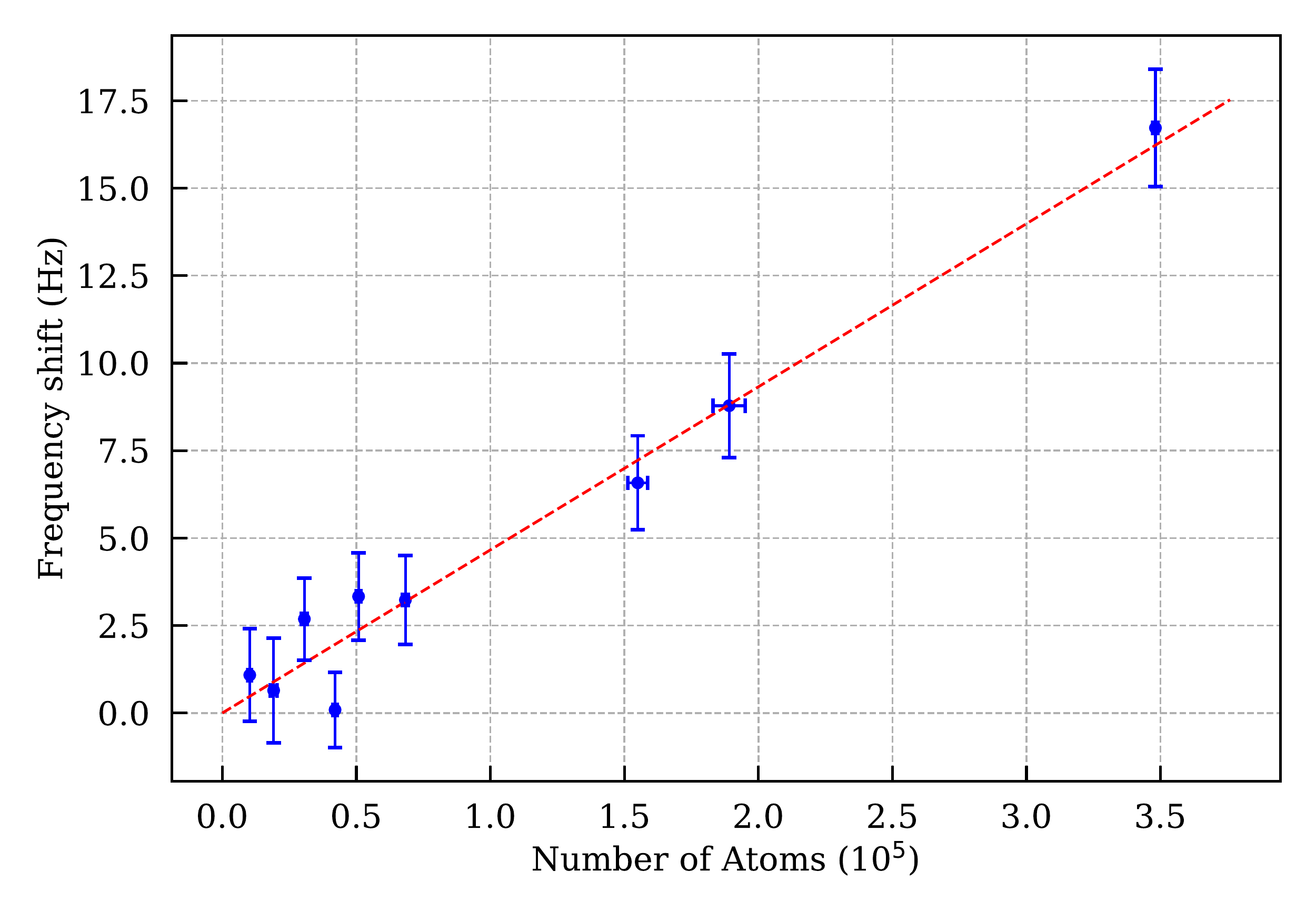}
    \caption{Evaluation of the density shift of the \srb clock transition. We assume linear scaling of density with atom number. }
    \label{fig:density_shift}
\end{figure}

Another important source of systematic uncertainty in bosonic \srb clocks is the density shift due to collisions~\cite{Lisdat2009}. We performed interleaved frequency shift measurements by changing the push power of our atomic source, thus tuning the lattice loading at will, as shown in Fig.~\ref{fig:density_shift}. Measurements spans between \num{3.5e5}  and \num{1e4} \srb atoms, while both the trap depth and the atomic temperature are kept constant. Narrow-line spectroscopy with $N = \num{4e4}$ atoms results in a frequency uncertainty of 0.2 Hz. The resulting density shift coefficient is

\begin{equation}
\Delta\nu_\rho = \bar{N}_\text{sites}\bar{V}_\text{site}\frac{\delta \nu}{N} = \num{7(2)e-18}\ \text{Hz m}^3,
\end{equation}
where $\bar{N}_\text{sites}$ is the number of occupied lattice sites, $\bar{V}_\text{site}$ is the average volume occupied by the atoms in each lattice {site. This} depends on the lattice depth $U_0$ and the atomic temperatures $T_r$ and $T_z$, as measured from {the} sideband spectroscopy (see Sec.~\ref{sec:SRS}) and calculated according to Ref.~\cite{Swallows12}. The   uncertainty  in the density coefficient is mainly due to the axial dimension estimation of the atomic sample {due to} the alignment of our imaging system. Compared with previous measurements~\cite{Lisdat2009}, we found a discrepancy of a factor 3 {for $\Delta\nu_\rho$}\footnote{We recalculated the coefficient from \cite{Lisdat2009} using the correct expressions for $\bar{N}_\text{sites}$ and $\bar{V}_\text{site}$ according to \cite{Swallows12}. The new value is \num{2.2e-17} Hz m$^3$ assuming a temperature of 4 $\mu$K. The discrepancy can be eliminated assuming a temperature of 2 $\mu$K.}.

Another environmental source of frequency shift is the blackbody radiation (BBR) due to the surrounding  ambient temperature $T$~\cite{Porsev2006}. An accurate evaluation of this effect is beyond the scope of this work. We point out that the MOT/bias coils are not thermally controlled, but due to the low duty cycle their effect is negligible compared to thermal fluctuations of the laboratory, which are currently controlled only within 0.5 K. The projected uncertainty in the BBR shift is \num{2e-17} in relative units, which is completely negligible at the current level of  accuracy.

Finally, all the frequency chain driving the AOMs used for clock spectroscopy is referenced to a H-maser {used for the realization of} UTC(IT) timescale. The estimated uncertainty due to residual phase noise after phase-locked loop to the external reference is evaluated below 0.3 Hz at 1 s. This results in an uncertainty of \num{1e-2} Hz for typical interleaved clock averaging times. 

{In summary}, the total  frequency accuracy is  \SI{8.4}{Hz}, or \num{2e-14} in relative units. It is mainly limited by the statistical uncertainty  in the determination of the quadratic Zeeman shift and the probe and lattice Stark shifts.

\section{Conclusion}
We have described a novel apparatus for a Sr optical lattice clock based on an optically-controlled cold atomic source and a multi-wavelength frequency stabilization system. These ingredients enable long periods of operation with low maintenance and high stability. Thanks to our 2D-MOT based atomic source, we have demonstrated a lifetime  of \SI{18.2(8)}{s} in the optical lattice{, remarkably longer than those obtained with Zeeman slowers.} 

A single reference cavity is able to properly  stabilize all the {lasers}, resulting in a{n} efficient and compact stabilization unit. 
{The multi-wavelength stabilization method} offers an ultimate limit on the clock laser instability at the level of \SI{1.1(4)e-16 }{ { / }\sqrt{\tau}} without any laser amplitude stabilization.

The resulting system is simpler and more cost effective than previous realizations, making this design suitable for applications in challenging real-world environments~\cite{Takamoto2020} and for industrial grade system.

Our optical clock  has shown a long-term frequency instability as low as $\num{4e-14}~/\sqrt{\tau}$, estimated by means of interleaved clock operation. Such limited stability is primarily affected by the short local oscillator coherence time {caused by} environmental vibration noise. Stable operation dominated by white frequency noise has been proved by interleaved frequency measurements for measurement times of { several hours}.

Technical frequency shifts have been resolved with less than \SI{10}{Hz}  uncertainty, limited by the short coherence time of the local oscillator and by the relatively short averaging time. 

{ The system described in this work offers much room for improvements, both regarding the short-term clock stability and the long-term clock operation. The short-term stability would immediately benefit from adding a commercially-available anti-vibration system for the reference cavity and by reducing the cycle dead-time by implementing a parallel data processing of the CCD data. These two simple changes should lower the clock instability by more than a factor 10. Long-term clock operation would require an active compensation of the multi-wavelength drift affecting the lattice loading process and the addition of automated locking algorithms for all the necessary lasers. Finally, a complete accuracy assessment of the system will require a better control of the room temperature and all clock components in order to correctly quantify the impact of the BBR shift.}

Immediate gain in frequency stability can  {  also} be obtained by optical frequency comb-assisted spectral purity transfer~\cite{Hagemann13} using {the} more stable optical local oscillator available for INRIM's Yb clock~\cite{Barbieri2019}. Fast and stable clock operations are also key ingredients to study quantum-enhanced technologies to be implemented in the newly designed science cell~\cite{Tarallo20}. The upgraded system can also test active generation of a frequency standard using its sideband-enhanced cold atomic beam~\cite{Liu2020}.




\section*{Acknowledgment}

We thank F. Bregolin { and G. Barontini} for careful reading of the manuscript, and M. Bober for useful discussions.
We acknowledge funding of the project EMPIR-USOQS; EMPIR projects are co-funded by the European Union’ Horizon 2020 research and innovation program and the EMPIR participating states. 
We also acknowledge funding from the QuantERA project Q-Clocks.




\bibliographystyle{IEEEtran}
\bibliography{paper.bib}

\begin{thebibliography}{10}
\providecommand{\url}[1]{#1}
\csname url@samestyle\endcsname
\providecommand{\newblock}{\relax}
\providecommand{\bibinfo}[2]{#2}
\providecommand{\BIBentrySTDinterwordspacing}{\spaceskip=0pt\relax}
\providecommand{\BIBentryALTinterwordstretchfactor}{4}
\providecommand{\BIBentryALTinterwordspacing}{\spaceskip=\fontdimen2\font plus
\BIBentryALTinterwordstretchfactor\fontdimen3\font minus
  \fontdimen4\font\relax}
\providecommand{\BIBforeignlanguage}[2]{{%
\expandafter\ifx\csname l@#1\endcsname\relax
\typeout{** WARNING: IEEEtran.bst: No hyphenation pattern has been}%
\typeout{** loaded for the language `#1'. Using the pattern for}%
\typeout{** the default language instead.}%
\else
\language=\csname l@#1\endcsname
\fi
#2}}
\providecommand{\BIBdecl}{\relax}
\BIBdecl

\bibitem{Campbell2017}
\BIBentryALTinterwordspacing
S.~L. Campbell, R.~B. Hutson, G.~E. Marti, A.~Goban, N.~D. Oppong, R.~L.
  McNally, L.~Sonderhouse, J.~M. Robinson, W.~Zhang, B.~J. Bloom, and J.~Ye,
  ``A fermi-degenerate three-dimensional optical lattice clock,''
  \emph{Science}, vol. 358, no. 6359, pp. 90--94, Oct. 2017. [Online].
  Available: \url{https://doi.org/10.1126/science.aam5538}
\BIBentrySTDinterwordspacing

\bibitem{Schioppo2017}
M.~Schioppo, R.~C. Brown, W.~F. McGrew, N.~Hinkley, R.~J. Fasano, K.~Beloy,
  T.~Yoon, G.~Milani, D.~Nicolodi, J.~Sherman \emph{et~al.}, ``Ultrastable
  optical clock with two cold-atom ensembles,'' \emph{Nature Photonics},
  vol.~11, no.~1, pp. 48--52, 2017.

\bibitem{Beloy2021}
\BIBentryALTinterwordspacing
K.~Beloy, M.~I. Bodine, T.~Bothwell, S.~M. Brewer, S.~L. Bromley, J.-S. Chen,
  J.-D. Deschênes, S.~A. Diddams, R.~J. Fasano, T.~M. Fortier, Y.~S. Hassan,
  D.~B. Hume, D.~Kedar, C.~J. Kennedy, I.~Khader, A.~Koepke, D.~R. Leibrandt,
  H.~Leopardi, A.~D. Ludlow, W.~F. McGrew, W.~R. Milner, N.~R. Newbury,
  D.~Nicolodi, E.~Oelker, T.~E. Parker, J.~M. Robinson, S.~Romisch, S.~A.
  Schäffer, J.~A. Sherman, L.~C. Sinclair, L.~Sonderhouse, W.~C. Swann,
  J.~Yao, J.~Ye, X.~Zhang, and B.~A. C. O. N.~B. Collaboration*, ``Frequency
  ratio measurements at 18-digit accuracy using an optical clock network,''
  \emph{Nature}, vol. 591, no. 7851, pp. 564--569, 2021. [Online]. Available:
  \url{https://doi.org/10.1038/s41586-021-03253-4}
\BIBentrySTDinterwordspacing

\bibitem{Riehle2015}
\BIBentryALTinterwordspacing
F.~Riehle, ``Towards a redefinition of the second based on optical atomic
  clocks,'' \emph{Comptes Rendus Physique}, vol.~16, no.~5, pp. 506--515, 2015,
  the measurement of time / La mesure du temps. [Online]. Available:
  \url{https://www.sciencedirect.com/science/article/pii/S1631070515000638}
\BIBentrySTDinterwordspacing

\bibitem{Ushijima2015}
\BIBentryALTinterwordspacing
I.~Ushijima, M.~Takamoto, M.~Das, T.~Ohkubo, and H.~Katori, ``Cryogenic optical
  lattice clocks,'' \emph{Nature Photonics}, vol.~9, no.~3, pp. 185--189, Feb.
  2015. [Online]. Available: \url{https://doi.org/10.1038/nphoton.2015.5}
\BIBentrySTDinterwordspacing

\bibitem{Schwarz2020}
\BIBentryALTinterwordspacing
R.~Schwarz, S.~D\"{o}rscher, A.~Al-Masoudi, E.~Benkler, T.~Legero, U.~Sterr,
  S.~Weyers, J.~Rahm, B.~Lipphardt, and C.~Lisdat, ``Long term measurement of
  the sr87 clock frequency at the limit of primary cs clocks,'' \emph{Physical
  Review Research}, vol.~2, no.~3, Aug. 2020. [Online]. Available:
  \url{https://doi.org/10.1103/physrevresearch.2.033242}
\BIBentrySTDinterwordspacing

\bibitem{Madjarov2019}
\BIBentryALTinterwordspacing
I.~S. Madjarov, A.~Cooper, A.~L. Shaw, J.~P. Covey, V.~Schkolnik, T.~H. Yoon,
  J.~R. Williams, and M.~Endres, ``An atomic-array optical clock with
  single-atom readout,'' \emph{Physical Review X}, vol.~9, no.~4, Dec. 2019.
  [Online]. Available: \url{https://doi.org/10.1103/physrevx.9.041052}
\BIBentrySTDinterwordspacing

\bibitem{Young2020}
\BIBentryALTinterwordspacing
A.~W. Young, W.~J. Eckner, W.~R. Milner, D.~Kedar, M.~A. Norcia, E.~Oelker,
  N.~Schine, J.~Ye, and A.~M. Kaufman, ``Half-minute-scale atomic coherence and
  high relative stability in a tweezer clock,'' \emph{Nature}, vol. 588, no.
  7838, pp. 408--413, Dec. 2020. [Online]. Available:
  \url{https://doi.org/10.1038/s41586-020-3009-y}
\BIBentrySTDinterwordspacing

\bibitem{Miyake2019}
\BIBentryALTinterwordspacing
H.~Miyake, N.~C. Pisenti, P.~K. Elgee, A.~Sitaram, and G.~K. Campbell,
  ``Isotope-shift spectroscopy of the s01$\rightarrow$p13 and
  s01$\rightarrow$p03 transitions in strontium,'' \emph{Physical Review
  Research}, vol.~1, no.~3, Nov. 2019. [Online]. Available:
  \url{https://doi.org/10.1103/physrevresearch.1.033113}
\BIBentrySTDinterwordspacing

\bibitem{Grotti2018}
\BIBentryALTinterwordspacing
J.~Grotti, S.~Koller, S.~Vogt, S.~H\"{a}fner, U.~Sterr, C.~Lisdat, H.~Denker,
  C.~Voigt, L.~Timmen, A.~Rolland, F.~N. Baynes, H.~S. Margolis, M.~Zampaolo,
  P.~Thoumany, M.~Pizzocaro, B.~Rauf, F.~Bregolin, A.~Tampellini, P.~Barbieri,
  M.~Zucco, G.~A. Costanzo, C.~Clivati, F.~Levi, and D.~Calonico, ``Geodesy and
  metrology with a transportable optical clock,'' \emph{Nature Physics},
  vol.~14, no.~5, pp. 437--441, Feb. 2018. [Online]. Available:
  \url{https://doi.org/10.1038/s41567-017-0042-3}
\BIBentrySTDinterwordspacing

\bibitem{Takamoto2020}
\BIBentryALTinterwordspacing
M.~Takamoto, I.~Ushijima, N.~Ohmae, T.~Yahagi, K.~Kokado, H.~Shinkai, and
  H.~Katori, ``Test of general relativity by a pair of transportable optical
  lattice clocks,'' \emph{Nature Photonics}, vol.~14, no.~7, pp. 411--415, Apr.
  2020. [Online]. Available: \url{https://doi.org/10.1038/s41566-020-0619-8}
\BIBentrySTDinterwordspacing

\bibitem{Bothwell2021}
T.~Bothwell, C.~J. Kennedy, A.~Aeppli, D.~Kedar, J.~M. Robinson, E.~Oelker,
  A.~Staron, and J.~Ye, ``Resolving the gravitational redshift within a
  millimeter atomic sample,'' 2021.

\bibitem{Nevsky:13}
\BIBentryALTinterwordspacing
A.~Nevsky, S.~Alighanbari, Q.-F. Chen, I.~Ernsting, S.~Vasilyev, S.~Schiller,
  G.~Barwood, P.~Gill, N.~Poli, and G.~M. Tino, ``Robust frequency
  stabilization of multiple spectroscopy lasers with large and tunable offset
  frequencies,'' \emph{Opt. Lett.}, vol.~38, no.~22, pp. 4903--4906, Nov 2013.
  [Online]. Available:
  \url{http://www.osapublishing.org/ol/abstract.cfm?URI=ol-38-22-4903}
\BIBentrySTDinterwordspacing

\bibitem{Ohmae2021}
N.~Ohmae, M.~Takamoto, Y.~Takahashi, M.~Kokubun, K.~Araki, A.~Hinton,
  I.~Ushijima, T.~Muramatsu, T.~Furumiya, Y.~Sakai, N.~Moriya, N.~Kamiya,
  K.~Fujii, R.~Muramatsu, T.~Shiimado, and H.~Katori, ``Transportable strontium
  optical lattice clocks operated outside laboratory at the level of 10 -18
  uncertainty,'' \emph{Advanced Quantum Technologies}, p. 2100015, may 2021.

\bibitem{Milani2017}
G.~Milani, B.~Rauf, P.~Barbieri, F.~Bregolin, M.~Pizzocaro, P.~Thoumany,
  F.~Levi, and D.~Calonico, ``Multiple wavelength stabilization on a single
  optical cavity using the offset sideband locking technique,'' \emph{Optics
  Letters}, vol.~42, no.~10, p. 1970, may 2017.

\bibitem{Gibble13}
\BIBentryALTinterwordspacing
K.~Gibble, ``Scattering of cold-atom coherences by hot atoms: Frequency shifts
  from background-gas collisions,'' \emph{Phys. Rev. Lett.}, vol. 110, p.
  180802, May 2013. [Online]. Available:
  \url{https://link.aps.org/doi/10.1103/PhysRevLett.110.180802}
\BIBentrySTDinterwordspacing

\bibitem{Barbiero2020}
\BIBentryALTinterwordspacing
M.~Barbiero, M.~G. Tarallo, D.~Calonico, F.~Levi, G.~Lamporesi, and G.~Ferrari,
  ``Sideband-enhanced cold atomic source for optical clocks,'' \emph{Physical
  Review Applied}, vol.~13, no.~1, Jan. 2020. [Online]. Available:
  \url{https://doi.org/10.1103/physrevapplied.13.014013}
\BIBentrySTDinterwordspacing

\bibitem{Taich2006}
\BIBentryALTinterwordspacing
A.~V. Taichenachev, V.~I. Yudin, C.~W. Oates, C.~W. Hoyt, Z.~W. Barber, and
  L.~Hollberg, ``Magnetic field-induced spectroscopy of forbidden optical
  transitions with application to lattice-based optical atomic clocks,''
  \emph{Phys. Rev. Lett.}, vol.~96, p. 083001, Mar 2006. [Online]. Available:
  \url{https://link.aps.org/doi/10.1103/PhysRevLett.96.083001}
\BIBentrySTDinterwordspacing

\bibitem{Takamoto2005}
M.~Takamoto, F.-L. Hong, R.~Higashi, and H.~Katori, ``An optical lattice
  clock,'' \emph{Nature}, vol. 435, no. 7040, pp. 321--324, may 2005.

\bibitem{Corning}
\BIBentryALTinterwordspacing
Corning, ``See corning technical broshure.'' [Online]. Available:
  \url{www.corning.com/ule.}
\BIBentrySTDinterwordspacing

\bibitem{Thorpe2008}
J.~I. Thorpe, K.~Numata, and J.~Livas, ``Laser frequency stabilization and
  control through offset sideband locking to optical cavities,'' \emph{Optics
  Express}, vol.~16, no.~20, p. 15980, sep 2008.

\bibitem{Akatsuka08}
\BIBentryALTinterwordspacing
T.~Akatsuka, M.~Takamoto, and H.~Katori, ``Optical lattice clocks with
  non-interacting bosons and fermions,'' \emph{Nature Physics}, vol.~4, no.~12,
  pp. 954--959, 2008. [Online]. Available:
  \url{https://doi.org/10.1038/nphys1108}
\BIBentrySTDinterwordspacing

\bibitem{Haefner2015}
S.~Häfner, S.~Falke, C.~Grebing, S.~Vogt, T.~Legero, M.~Merimaa, C.~Lisdat,
  and U.~Sterr,
  ``8{\hspace{0.167em}}{\hspace{0.167em}}{\texttimes}{\hspace{0.167em}}{\hspace{0.167em}}10{\^{}}-17
  fractional laser frequency instability with a long room-temperature cavity,''
  \emph{Optics Letters}, vol.~40, no.~9, p. 2112, may 2015.

\bibitem{Katori1999}
\BIBentryALTinterwordspacing
H.~Katori, T.~Ido, Y.~Isoya, and M.~Kuwata-Gonokami, ``Magneto-optical trapping
  and cooling of strontium atoms down to the photon recoil temperature,''
  \emph{Phys. Rev. Lett.}, vol.~82, pp. 1116--1119, Feb 1999. [Online].
  Available: \url{https://link.aps.org/doi/10.1103/PhysRevLett.82.1116}
\BIBentrySTDinterwordspacing

\bibitem{Hall1999}
J.~L. Hall, M.~S. Taubman, and J.~Ye, ``Laser stabilization,'' in
  \emph{HANDBOOK OF OPTICS}, Bass, Ed.\hskip 1em plus 0.5em minus 0.4em\relax
  New York: McGraw-Hill, 2001, vol.~4, ch.~27.

\bibitem{Nazarova_2006}
T.~Nazarova, F.~Riehle, and U.~Sterr, ``Vibration-insensitive reference cavity
  for an ultra-narrow-linewidth laser,'' \emph{Applied Physics B}, vol.~83,
  no.~4, pp. 531--536, may 2006.

\bibitem{Fiori2004}
\BIBentryALTinterwordspacing
F.~Acernese, P.~Amico, N.~Arnaud, D.~Babusci, R.~Barill{\'{e}}, F.~Barone,
  L.~Barsotti, M.~Barsuglia, F.~Beauville, M.~A. Bizouard, C.~Boccara,
  F.~Bondu, L.~Bosi, C.~Bradaschia, L.~Bracci, S.~Braccini, A.~Brillet,
  V.~Brisson, L.~Brocco, D.~Buskulic, G.~Calamai, E.~Calloni, E.~Campagna,
  F.~Cavalier, G.~Cella, E.~Chassande-Mottin, F.~Cleva, T.~Cokelaer, C.~Corda,
  J.~P. Coulon, E.~Cuoco, V.~Dattilo, M.~Davier, R.~D. Rosa, L.~D. Fiore, A.~D.
  Virgilio, B.~Dujardin, A.~Eleuteri, D.~Enard, I.~Ferrante, F.~Fidecaro,
  I.~Fiori, R.~Flaminio, J.~D. Fournier, S.~Frasca, F.~Frasconi, L.~Gammaitoni,
  A.~Gennai, A.~Giazotto, G.~Giordano, G.~Guidi, H.~Heitmann, P.~Hello,
  P.~Heusse, L.~Holloway, S.~Kreckelbergh, P.~L. Penna, V.~Loriette,
  M.~Loupias, G.~Losurdo, J.~M. Mackowski, E.~Majorana, C.~N. Man, F.~Marion,
  F.~Martelli, A.~Masserot, L.~Massonnet, M.~Mazzoni, L.~Milano, J.~Moreau,
  F.~Moreau, N.~Morgado, F.~Mornet, B.~Mours, J.~Pacheco, A.~Pai, C.~Palomba,
  F.~Paoletti, R.~Passaquieti, D.~Passuello, B.~Perniola, L.~Pinard,
  R.~Poggiani, M.~Punturo, P.~Puppo, K.~Qipiani, J.~Ramonet, P.~Rapagnani,
  V.~Reita, A.~Remillieux, F.~Ricci, I.~Ricciardi, G.~Russo, S.~Solimeno,
  R.~Stanga, A.~Toncelli, M.~Tonelli, E.~Tournefier, F.~Travasso, H.~Trinquet,
  M.~Varvella, D.~Verkindt, F.~Vetrano, O.~Veziant, A.~Vicer{\'{e}}, J.~Y.
  Vinet, H.~Vocca, and M.~Yvert, ``Properties of seismic noise at the virgo
  site,'' \emph{Class. Quantum Grav.}, vol.~21, no.~5, pp. S433--S440, feb
  2004. [Online]. Available: \url{https://doi.org/10.1088/0264-9381/21/5/008}
\BIBentrySTDinterwordspacing

\bibitem{Tarallo2011}
M.~G. Tarallo, N.~Poli, M.~Schioppo, D.~Sutyrin, and G.~Tino, ``A
  high-stability semiconductor laser system for a 88 sr-based optical lattice
  clock,'' \emph{Applied Physics B}, vol. 103, no.~1, pp. 17--25, 2011.

\bibitem{Barbiero2021}
M.~Barbiero, M.~G. Tarallo, F.~Rullo, M.~Risaro, C.~Clivati, D.~Calonico, and
  F.~Levi, ``Inrim sr optical clock: an optically loaded apparatus for
  high-stability metrology,'' in \emph{Proceedings of the ``2021 Joint
  Conference of the European Frequency and Time Forum and IEEE International
  Frequency Control Symposium (EFTF/IFCS)''}, 2021, p. 7233.

\bibitem{Bergquist1992}
J.~Bergquist, W.~Itano, and D.~Wineland, ``Laser stabilization to a single
  ion,'' in \emph{Frontiers in Laser Spectroscopy}, T.~Hansch and M.~Inguscio,
  Eds.\hskip 1em plus 0.5em minus 0.4em\relax North Holland, 1992, pp.
  359--376.

\bibitem{Quessada_2003}
A.~Quessada, R.~P. Kovacich, I.~Courtillot, A.~Clairon, G.~Santarelli, and
  P.~Lemonde, ``The dick effect for an optical frequency standard,''
  \emph{Journal of Optics B: Quantum and Semiclassical Optics}, vol.~5, no.~2,
  pp. S150--S154, apr 2003.

\bibitem{Grimm1999}
R.~Grimm, M.~Weidemüller, and Y.~B. Ovchinnikov, ``Optical dipole traps for
  neutral atoms,'' \emph{Advances in Atomic, Molecular and Optical Physics},
  vol.~42, pp. 95--170, 2000.

\bibitem{Drscher2018}
\BIBentryALTinterwordspacing
S.~D\"{o}rscher, R.~Schwarz, A.~Al-Masoudi, S.~Falke, U.~Sterr, and C.~Lisdat,
  ``Lattice-induced photon scattering in an optical lattice clock,''
  \emph{Physical Review A}, vol.~97, no.~6, Jun. 2018. [Online]. Available:
  \url{https://doi.org/10.1103/physreva.97.063419}
\BIBentrySTDinterwordspacing

\bibitem{Barbiero2019}
\BIBentryALTinterwordspacing
M.~Barbiero, ``Novel techniques for a strontium optical lattice clock,'' Ph.D.
  dissertation, Politecnico of Turin, 2019. [Online]. Available:
  \url{http://hdl.handle.net/11583/2750550}
\BIBentrySTDinterwordspacing

\bibitem{Blatt2009}
\BIBentryALTinterwordspacing
S.~Blatt, J.~W. Thomsen, G.~K. Campbell, A.~D. Ludlow, M.~D. Swallows, M.~J.
  Martin, M.~M. Boyd, and J.~Ye, ``Rabi spectroscopy and excitation
  inhomogeneity in a one-dimensional optical lattice clock,'' \emph{Phys. Rev.
  A}, vol.~80, p. 052703, Nov 2009. [Online]. Available:
  \url{https://link.aps.org/doi/10.1103/PhysRevA.80.052703}
\BIBentrySTDinterwordspacing

\bibitem{Pizzocaro2012}
M.~Pizzocaro, G.~A. Costanzo, A.~Godone, F.~Levi, A.~Mura, M.~Zoppi, and
  D.~Calonico, ``Realization of an ultrastable 578-nm laser for an yb lattice
  clock,'' \emph{IEEE transactions on ultrasonics, ferroelectrics, and
  frequency control}, vol.~59, no.~3, pp. 426--431, 2012.

\bibitem{Nicholson2015}
T.~L. Nicholson, S.~L. Campbell, R.~B. Hutson, G.~E. Marti, B.~J. Bloom, R.~L.
  McNally, W.~Zhang, M.~D. Barrett, M.~S. Safronova, G.~Strouse, W.~L. Tew, and
  J.~Ye, ``Systematic evaluation of an atomic clock at 2 $\times$ 10-18 total
  uncertainty,'' \emph{Nature Communications}, vol.~6, 2015.

\bibitem{Lisdat2009}
C.~Lisdat, J.~S. R.~V. Winfred, T.~Middelmann, F.~Riehle, and U.~Sterr,
  ``Collisional losses, decoherence, and frequency shifts in optical lattice
  clocks with bosons,'' \emph{Physical Review Letters}, vol. 103, no.~9, p.
  090801, aug 2009.

\bibitem{Swallows12}
M.~D. Swallows, M.~J. Martin, M.~Bishof, C.~Benko, Y.~Lin, S.~Blatt, A.~M. Rey,
  and J.~Ye, ``{Operating a 87Sr optical lattice clock with high precision and
  at high density},'' \emph{IEEE Transactions on Ultrasonics, Ferroelectrics,
  and Frequency Control}, vol.~59, no.~3, pp. 416--425, 2012.

\bibitem{Note1}
We recalculated the coefficient from \cite {Lisdat2009} using the correct
  expressions for $\bar {N}_\protect \text {sites}$ and $\bar {V}_\protect
  \text {site}$ according to \cite {Swallows12}. The new value is \num
  {2.2e-17} Hz m$^3$ assuming a temperature of 4 $\mu $K. The discrepancy can
  be eliminated assuming a temperature of 2 $\mu $K.

\bibitem{Porsev2006}
S.~G. Porsev and A.~Derevianko, ``Multipolar theory of blackbody radiation
  shift of atomic energy levels and its implications for optical lattice
  clocks,'' \emph{Physical Review A}, vol.~74, no.~2, p. 020502, 2006.

\bibitem{Hagemann13}
C.~Hagemann, C.~Grebing, T.~Kessler, S.~Falke, N.~Lemke, C.~Lisdat, H.~Schnatz,
  F.~Riehle, and U.~Sterr, ``Providing $10^{-16}$ short-term stability of a
  $1.5\mu\hbox{m}$ laser to optical clocks,'' \emph{IEEE Transactions on
  Instrumentation and Measurement}, vol.~62, no.~6, pp. 1556--1562, 2013.

\bibitem{Barbieri2019}
\BIBentryALTinterwordspacing
P.~Barbieri, C.~Clivati, M.~Pizzocaro, F.~Levi, and D.~Calonico, ``Spectral
  purity transfer with 5 {\texttimes} 10-17 instability at 1 s using a
  multibranch er:fiber frequency comb,'' \emph{Metrologia}, vol.~56, no.~4, p.
  045008, jul 2019. [Online]. Available:
  \url{https://doi.org/10.1088/1681-7575/ab2b0f}
\BIBentrySTDinterwordspacing

\bibitem{Tarallo20}
\BIBentryALTinterwordspacing
M.~G. Tarallo, ``Toward a quantum-enhanced strontium optical lattice clock at
  inrim,'' \emph{EPJ Web of Conferences}, vol. 230, p. 00011, 2020. [Online].
  Available: \url{http://dx.doi.org/10.1051/epjconf/202023000011}
\BIBentrySTDinterwordspacing

\bibitem{Liu2020}
\BIBentryALTinterwordspacing
H.~Liu, S.~B. J\"ager, X.~Yu, S.~Touzard, A.~Shankar, M.~J. Holland, and T.~L.
  Nicholson, ``Rugged mhz-linewidth superradiant laser driven by a hot atomic
  beam,'' \emph{Phys. Rev. Lett.}, vol. 125, p. 253602, Dec 2020. [Online].
  Available: \url{https://link.aps.org/doi/10.1103/PhysRevLett.125.253602}
\BIBentrySTDinterwordspacing

\end{thebibliography}
\end{document}